\documentstyle[epsfig,psfig]{texas}

\begin{document}

\title{Relation between Kilohertz QPOs and Inferred Mass Accretion
Rate in 4 LMXBs}

\author{Mariano M\'endez$^{1,2}$}
\address{(1) Astronomical Institute ``Anton Pannekoek'', University of
Amsterdam, Kruislaan 403, NL-1098 SJ Amsterdam, the Netherlands}
\address{(2) Facultad de Ciencias Astronomicas y Geofisicas, Universidad
Nacional de La Plata, Paseo del Bosque S/N, 1900 La Plata, Argentina\\
{\rm Email: mariano@astro.uva.nl}}

\begin{abstract}

I summarize the available {\em RXTE} data of the 4 low-mass X-ray
binaries (LMXBs) and {\em Atoll} sources Aql\,X--1, 4U\,1728--34,
4U\,1608--52, and 4U\,1636-53.  I concentrate on the relation between
the frequencies of the quasi-periodic oscillations at kilohertz
frequencies (kHz QPOs) and the X-ray flux and colors of these sources.
In these 4 sources the kHz QPOs are only observed in a narrow range of
spectral states (as defined from the X-ray color-color diagrams).  I
show that despite its complex dependence upon the X-ray flux, the
frequency of the kHz QPOs is monotonically related to the position of
the source in the color-color diagram.  These findings strengthen the
idea that in LMXBs the X-ray flux is not a good indicator of the mass
accretion rate, $\dot M$, and that the observed changes in the frequency
of the kHz QPOs in LMXBs are driven by changes in $\dot M$.

\end{abstract}

\section{Introduction}

In the past 3 years the Rossi X-ray Timing Explorer ({\em RXTE}) has
discovered kilohertz quasi-periodic oscillations (kHz QPOs) in the
persistent flux of 19 low-mass X-ray binaries (LMXB; see
\cite{vanderklis98} for a review).  In almost all cases the power
density spectra of these sources show twin kHz peaks that, as a function
of time, gradually move up and down in frequency, typically over a range
of several hundred Hz.

Initially, data from various sources seemed to indicate that the
separation $\Delta \nu$ between the twin kHz peaks remained constant
even as the peaks moved up and down in frequency \cite{strohmayer96a}.
In some sources a third, nearly-coherent, oscillation has been detected
during type-I X-ray bursts, at a frequency close to $\Delta \nu$, or
twice that value (see \cite{ssz98} for a review).  These two results
suggested that a beat-frequency mechanism was at work
\cite{strohmayer96a,miller98}, with the third peak being close to the
neutron star spin frequency or twice that.  In sources for which only
the twin kHz QPO, and no burst oscillations, were observed the frequency
difference was interpreted in terms of the neutron star spin frequency
as well.  But the simple beat-frequency interpretation of the kHz QPOs
is not without problems
\cite{vanderklisetal97a,mendez98b,mendez98a,ford1735,mendez1728}, and
other ideas discarding one or more elements of this basic picture, but
still predicting definite relations between the observed frequencies,
have been proposed \cite{titarchuk98,stella99,lamb99}.

One interesting result obtained from these {\em RXTE} observations is
the complex dependence of the QPO frequencies upon X-ray flux, which is
usually assumed to be a measure of the mass accretion rate $\dot M$.
One example is 4U\,1608--52 \cite{mendez99}:  While on time scales of
hours frequency and X-ray flux are well correlated, at epochs separated
by days to weeks the QPOs span the same range of frequencies even if the
average flux differs by a factor of 3 or more (see also
\cite{ford97a,zhangAqlx1}).  In this case, however, the QPO frequency is
very well correlated to the position of the source in the color-color
diagram \cite{mendez99}.

\begin{figure}[ht]
\centering
\epsfig{figure=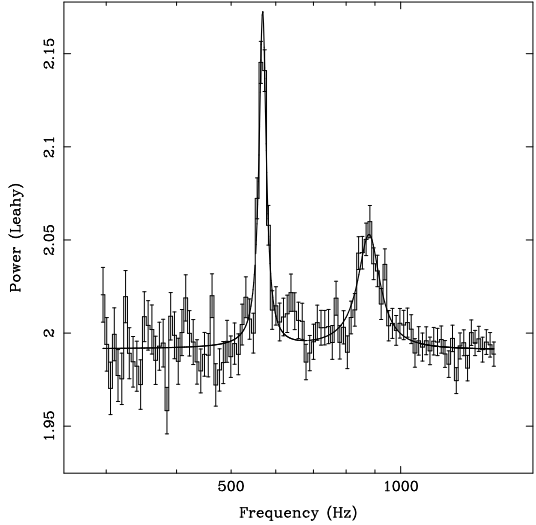, width=7.0cm, clip=}
\caption{Power spectrum of 4U\,1608--52 showing the two simultaneous kHz
QPOs.  This power spectrum was produced using $\sim 4200$ s of data from
1998 March 25 starting at 16:16 UTC.  On 64 s time scales the FWHM of
the QPO at $\sim 580$ Hz is $\sim 5$ Hz, but due to the variation of
its central frequency during the observation it appears broader in this
power spectrum.  The QPO at $\sim 900$ Hz is broader, and its FWHM on
time scales of 64 s is $\sim 100$ Hz.}
\label{figps}
\end{figure}

Here I summarize some results from {\em RXTE} observations of 4 LMXBs:
Aql\,X--1, 4U\,1728--34, 4U\,1608--52, and 4U\,1636-53.  (Some of these
results have been published before \cite{zhangAqlx1,mendez99,mendez1728}
or will be presented in more detail elsewhere
\cite{reigAqlx1,mendez1636}).  Here I focus on the relation of the
frequencies of the kHz QPOs to the X-ray flux and colors of the source.
These results cast some doubt about the recently reported detection of
the orbital frequency at the innermost stable orbit in 4U\,1820--30
\cite{zhang1820}.

\begin{figure}[t]
\centering
\epsfig{figure=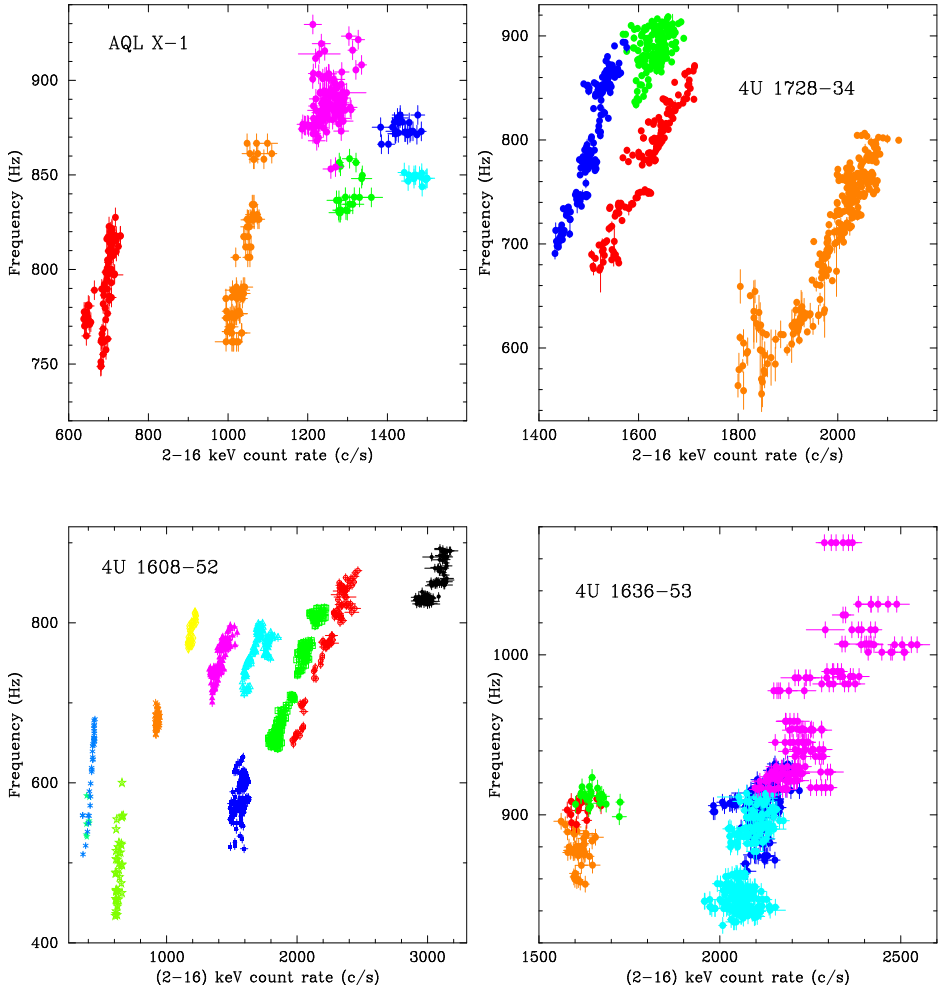, width=12.5cm, clip=}
\caption{Frequency of one of the kHz QPOs vs.  count rate for
Aql\,X--1, 4U\,1728--34, 4U\,1608--52, and 4U\,1636-53.  Different
symbols/colors indicate observations obtained on different dates.}
\label{figrate}
\end{figure}

\section{Results}

As an example, in Figure~\ref{figps} I show a power spectrum of
4U\,1608--52 in the range $300-1200$ Hz, where the two QPOs are seen
simultaneously.  Two, sometimes simultaneous, kHz QPOs are also present
in the power spectra of 4U\,1728--34 and 4U\,1636--53.  For Aql\,X--1
only one kHz QPO has been observed so far.

\begin{figure}[t]
\centering
\epsfig{figure=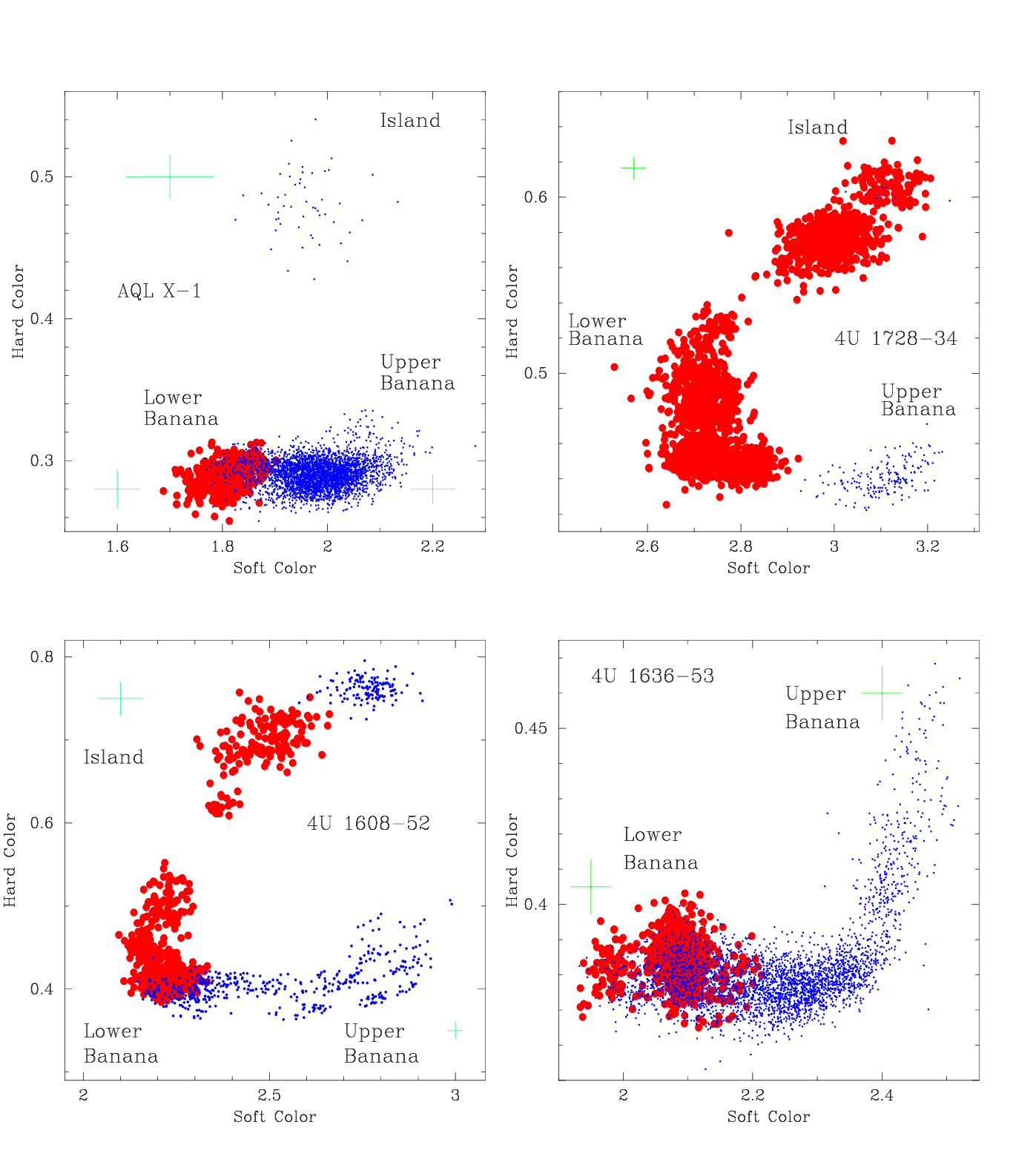, width=12.5cm, clip=}
\caption{Color-color diagrams of Aql\,X--1, 4U\,1728--34, 4U\,1608--52,
and 4U\,1636-53.  Typical error bars in different parts of the diagram
are shown.  Red circles and blue dots indicate segments with and without
kHz QPOs, respectively.}
\label{figcolor}
\end{figure}

As I already mentioned, the frequencies of these kHz QPOs slowly change
as a function of time.  In Figure~\ref{figrate} I show the relation of
the frequency of one of the kHz QPOs (for 4U\,1728--34, 4U\,1608--52,
and 4U\,1636--53 it is the kHz QPO at lower frequencies; for Aql\,X--1
it is the only kHz QPO detected so far) vs.  source count rate in the
$2-16$ keV energy range.  From this figure it is apparent that the
dependence of the kHz QPO frequencies upon X-ray intensity is complex
(the same result is obtained if the $2-16$ keV source flux is used
instead of the count rate).

In Figure~\ref{figcolor} I show the color-color diagrams of Aql\,X--1,
4U\,1728--34, 4U\,1608--52, and 4U\,1636-53.  The soft and hard colors
are defined as $I_{\rm (3.5-6.4) keV}/I_{\rm (2.0-3.5) keV}$, and
$I_{\rm (9.7-16.0) keV}/I_{\rm (6.4-9.7) keV}$, respectively, where $I$
is the background subtracted source count rate for the indicated energy
range.  These color-color diagrams are typical of the so-called {\em
Atoll} sources \cite{hasinger89}.  Except for 4U\,1636-53, which {\em
RXTE} only observed in the banana branch, the other three sources move
across all the branches of the atoll.

Interestingly, there seems to be a close relation between the position
of the source in the color-color diagram and the appearance of kHz QPOs
in the power spectrum:  the QPOs are only detected in the lower banana
and the moderate island states, and disappear both in the upper banana
and in the extreme island states (red circles and blue dots indicate
time intervals with and without kHz QPOs, respectively).

\begin{figure}[ht]
\centering
\epsfig{figure=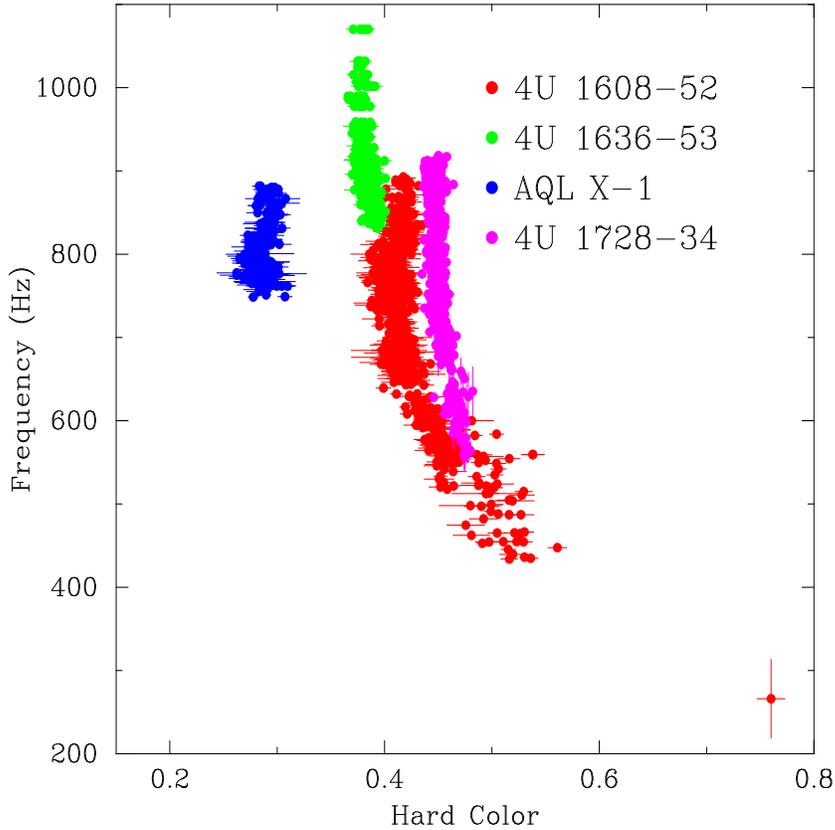, width=11.0cm, clip=}
\caption{The relation between the frequency of one of the kHz QPOs (for
4U\,1728--34, 4U\,1608--52, and 4U\,1636--53 it is the kHz QPO at lower
frequencies; for Aql\,X--1 it is the only kHz QPO detected so far) and
the hard color (see Fig.  \ref{figcolor}), for the same segments shown
in Figure \ref{figrate}.}
\label{fig_hc}
\end{figure}

In Figure~\ref{fig_hc} I present the relation of the frequency of one of
the kHz QPOs (as in Fig.~\ref{figrate}, for 4U\,1728--34, 4U\,1608--52,
and 4U\,1636--53 it is the kHz QPO at lower frequencies; for Aql\,X--1
it is the only kHz QPO detected so far) as a function of hard color (see
Fig.~\ref{figcolor}) for the same intervals shown in
Figure~\ref{figrate}.  The complexity seen in the frequency vs.  count
rate diagram (Fig.~\ref{figrate}) is reduced to a single track per
source in the frequency vs.  hard color diagram.

The shapes of the tracks in Figure~\ref{fig_hc} suggest that the hard
color may not be sensitive to changes of state when these sources move
into the banana in the color-color diagram (particularly in the case of
Aql\,X--1, because the banana branch is almost horizontal in this
diagram).  To further investigate this, I parametrized the color-color
diagram in terms of a one-dimensional variable that measures the
position of the source along the atoll.  I call this variable $S_{\rm
a}$, in analogy to what is usually called $S_{\rm Z}$ for the other
class of LMXBs, the {\em Z} sources.  The shape of the color-color
diagram is approximated with a spline, and a value of $S_{\rm a}$ is
assigned to each point according to the distance (along the spline) of
that point to a reference point in the diagram.

\begin{figure}[t]
\centering
\epsfig{figure=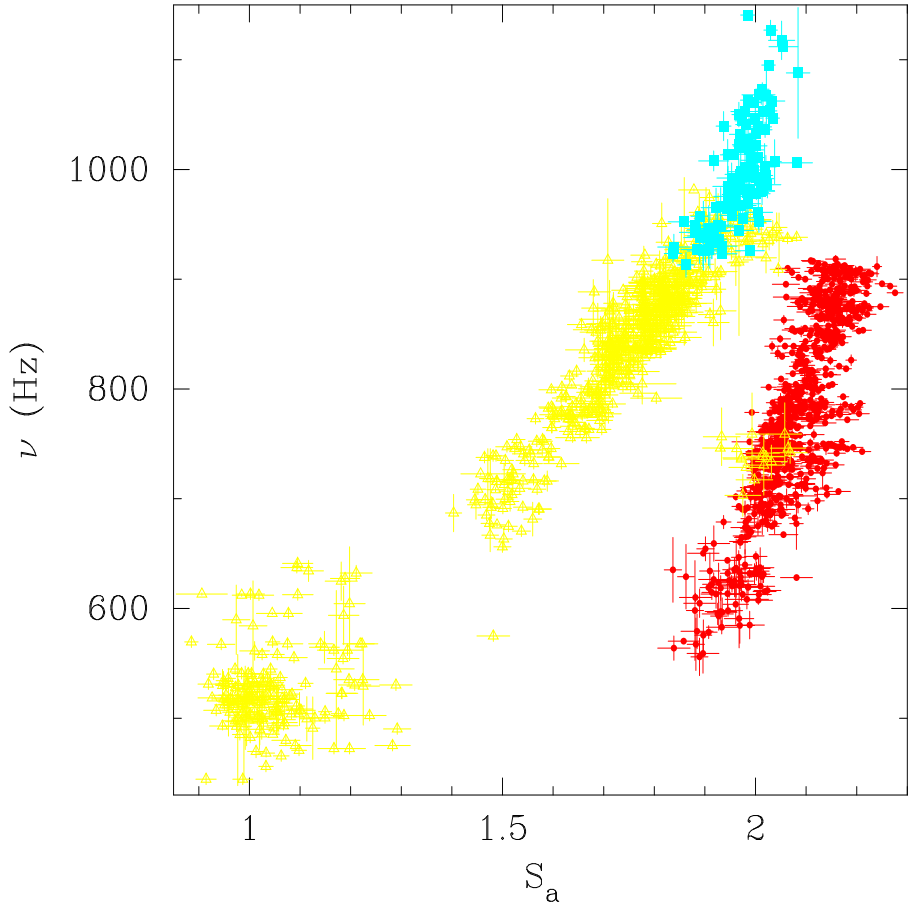, width=12.0cm, clip=}
\caption{Diagram of the frequencies of the kHz QPOs at lower and higher
frequencies vs.  the position of the source in the color-color diagram
for 4U\,1728--34, as measured by $S_{\rm a}$ (see Fig.~\ref{figcolor},
and text).  Red circles and blue squares represent the kHz peak at lower
and higher frequencies, respectively.  Yellow triangles represent
segments where I only detected one of the kHz QPOs in the power
spectrum.}
\label{figsa}
\end{figure}

In Figure~\ref{figsa} I present the relation between the frequencies of
the two QPOs in 4U\,1728--34 vs.  $S_{\rm a}$.  I arbitrarily defined
$S_{\rm a} = 1$ at colors (3.02,0.59), and $S_{\rm a} = 2$ at colors
(2.75,0.46) (see Fig.~\ref{figcolor}).  Red circles in this figure
correspond to the kHz QPO at lower frequencies (the same data presented
in Figure~\ref{figrate}).  The blue squares correspond to the frequency
of the kHz QPO at higher frequencies, and the yellow triangles
correspond to measurements in which I only detect one of the kHz QPOs;
however, from the location of each point in this diagram it is possible
to determine whether it is the QPO at higher or lower frequencies.

\section{Discussion}

These results show that a total lack of correlation between frequency
and count rate on time scales longer than a day (Fig.~\ref{figrate}) can
coexist with a very good correlation between frequency and position in
the X-ray color-color diagram (Fig.~\ref{fig_hc} and \ref{figsa}).  The
frequency of the QPO increases with $S_{\rm a}$, as the source moves
from the island to the banana.  Only on time scales of hours does the
QPO frequency appear to also correlate well with count rate.  The
presence of the QPOs also correlates well with the position in the
color-color diagram:  the QPOs are only detected in the lower banana and
the moderate island states, and disappear both in the upper banana and
in the extreme island states (Fig.~\ref{figcolor}).

In the {\em Atoll} sources $\dot M$ is thought to increase monotonically
with $S_{\rm a}$ along the track in the color-color diagram, from the
island to the upper banana \cite{hasinger89}, whereas X-ray count rate
tracks $\dot M$ much less well \cite{vdk90,vdk94apj,prins97}.  The
properties of the power spectra below $\sim 100$ Hz depend monotonically
upon $S_{\rm a}$ \cite{hasinger89}.  The result that the frequency of
the kHz QPO is well correlated to $S_{\rm a}$, but not to X-ray count
rate, implies that the kHz QPO frequency {\em also} depends
monotonically upon inferred $\dot M$.  In {\em Z} sources similar
conclusions have also been reached \cite{wijnands17+2}.

Further support for this interpretation comes from the simultaneous
analysis of the low and high frequency parts of the power spectra.  In
4U\,1728--34 the kHz QPO frequencies were recently found to be very well
correlated to several $< 100$ Hz power spectral properties
\cite{ford&vdk98}, while a similar result was obtained for a number of
other {\em Atoll} (and {\em Z}) sources \cite{psaltisetal98}.  In all
these sources, not only the position in the color-color diagram and the
various low frequency power spectral parameters, but also the
frequencies of the kHz QPOs are all well correlated with each other.
This indicates that the single parameter, inferred to be $\dot M$, which
governs all the former properties also governs the frequency of the kHz
QPO.

X-ray intensity is the exception:  it can vary more or less
independently from the other parameters.  In 4U\,1608--52, it can change
by a factor of $\sim 4$ (see Fig.~\ref{figrate}) while the other
parameters do not vary significantly.  If as inferred, this constancy of
the other parameters means that $\dot M$ does not change, then this
indicates that strongly variable beaming of the X-ray flux, or
large-scale redistribution of some of the radiation over unobserved
energy ranges is occurring in order to change the flux by the observed
factors, {\em without} any appreciable changes in the X-ray spectrum.
Perhaps the $\dot M$ governing all the other parameters is the $\dot M$
through the inner accretion disk, whereas matter also flows onto the
neutron star in a more radial inflow, or away from it in a jet.

All current models propose that the frequencies of the kHz QPOs, which
are thought to reflect the Keplerian orbital frequency of accreting
matter in a disk around the neutron star, increase monotonically with
$\dot M$, because the inner edge of the disk moves in when $\dot M$
increases.  However, the accretion disk cannot move closer to the
neutron star than the radius of the innermost stable circular orbit,
even if $\dot M$ keeps increasing.  This means that the frequency of the
kHz QPOs should ``saturate'' at a value corresponding to the Keplerian
frequency at the innermost stable orbit for a given source.

None of these 4 sources show evidence for a saturation of the frequency
of the kHz QPOs at a constant maximum value as $\dot M$ increases,
different from what was inferred for 4U\,1820--30 \cite{zhang1820}.  In
4U\,1820--30 the kHz QPO frequencies increase with count rate up to a
threshold level, above which the frequencies remain approximately
constant while the count rate keeps increasing.  Assuming that count
rate is a measure for $\dot M$, this was interpreted as evidence for the
inner edge of the disk reaching the general-relativistic innermost
stable orbit.  However, the results presented above indicate that count
rate is not a good measure for $\dot M$.  Inspection of
Figure~\ref{figrate} suggests that with sparser sampling some of those
plots could easily have looked similar to that of 4U\,1820--30.  It will
therefore be of great interest to see if in 4U\,1820--30 the saturation
of QPO frequency as a function of count rate is still there when this
parameter is plotted as a function of position in the X-ray color-color
diagram.

\section*{Acknowledgments}

This work was supported in part by the Netherlands Research School for
Astronomy (NOVA), the Leids Kerkhoven-Bosscha Fonds (LKBF), and the
NWO Spinoza grant 08-0 to E.P.J. van den Heuvel.  MM is a fellow of the
Consejo Nacional de Investigaciones Cientificas y Tecnicas de la
Republica Argentina.  This research has made use of data obtained
through the High Energy Astrophysics Science Archive Research Center
Online Service, provided by the NASA/Goddard Space Flight Center.

\end{document}